\documentclass[a4paper]{article} 

\usepackage{ai_fncs,eqn_center}
\begin{document}

\twocolumn[
\begin{center}
{\bf Microscopic Energy Current Field with Multi--body Force
 in Hamiltonian System}\\
~\\
{\bf \underline{Atsushi Ito}}$^\dagger$ and {\bf Hiroaki Nakamura$^\ddagger$}\\
$^\dagger$Department of physics, Graduate School of Science, 
Nagoya University, Chikusa, Nagoya 464--8602, Japan\\
$^\ddagger$National Institute for Fusion Science,
Oroshi--cho 322--6, Toki 509--5292, Japan\\~\\
(received~****)\\
\end{center}
]

\begin{sloppypar}
\begin{abstract}
{\bf
Microscopic energy current can be derived from 
 microscopic energy field localized only in particle positions.
If the energy current is applied to classical molecular dynamics,
 it is expected to become a new information.
However,
  multi--body force except two--body force causes obscurity
 when multi--body interaction energy is localized.
In the present study, our new method enables to determine
 the localization of the multi--body interaction energy.
We obtain the energy current between particles by the law of the multi--body force
 corresponding to Newton's third law on two--body force.
}

\noindent{\it Keywords} ; Energy current, Multi--body force, Molecular dynamics

\end{abstract}

\section{Introduction}


If a physical quantity and its flow exist, they need to be fields.
The energy should be a field if the energy current is considerd.
The total energy is, however,  one quantity in the phase space.
It is namely not the field.
To accept the existence of the energy current,
 microscopic energy field is defined
 which all sort of the energy is localized only in particle positions.
The energy current field is derived from the microscopic energy field.
S. Lepri, R. Livi and A. Politi decided the energy current 
 with the approximation of low--$k$ limit \cite{Lepri}.
S. Takesue derived the energy current field which satisfies strictly the `continuity equation'
 with the microscopic energy field \cite{Takesue}.
In his theory, the energy current field includes
 the term of the energy current between particles.
The his theory differs from the one of J. H. Irving and G. Kirkwood \cite{IrvingKirkwood}
 in the point that a distribution function is not required.

The microscopic energy field is not zero only at particle positions
 by the use of Dirac's delta function.
The kinetic energy of each particle is naturally localized 
 in the its particle position.
Moreover, not only the kinetic energy but also the interaction energy
 is localized in particle positions.
However, the interaction energy cannot be possessed by a single particle
 because the other party is necessary for the interaction.
Historically, it was thought that both of two particles
 doing two--body interaction have just half the interaction energy.

If the theory of the microscopic energy field is expanded
 into a classical molecular dynamics (CMD),
 the energy current between particles can become a new information
 to analyze the dynamics of the particle.
When particles are atoms and have covalent bonds,
 they are often subjected by the multi--body force in CMD.
For the energy current, the multi--body interaction energy should be localized in the particle positions.
The problem of the localization of the multi--body interaction energy is not
 as simple as the localization of the two--body interaction energy.

%
%

For example, let us consider a water molecule composed of two hydrogen atoms
 and an oxygen atom.
Three--body interaction is often adopted to keep the angle between two covalent bonds $109.4$ degree.
The three--body interaction potential function depends on three atom positions.
Therefore, the three--body interaction energy should be localized in the positions of the three atoms.
One means of localization is to set one--third of the three--body interaction energy on each particles.
However, we consider this means is not correct because the oxygen atom and the hydrogen atoms differ in their condition,
 both in the kind of atoms and the magnitude of the forces.



In another scene, such a problem of the localization the three--body interaction energy
 is discussed historically \cite{Chen,Volz,Hardy}.
They argue the three particles has one--third of the three--body interaction energy.
However they are groundless.
In addition, the problem of the localization of the interaction energy has not been discussed
 when it is greater than three--body interaction.

In the present paper, we propose new method of the localization of
 the multi--body interaction energy into particle positions.
The law of the multi--body force corresponding to Newton's third law is used.

The theory of the microscopic energy field and the energy current
 on a three dimensional space is introduced in section 2.
We derive the energy current between particles according to only two--body force in section 3.
In section 4, we describe the new method for the multi--body force.
And, we discuss our new method in section 5.


\section{Microscopic Energy Field\\and Energy Current}

We introduce the derivation of microscopic energy field
 and microscopic energy current on Takesue's work\cite{Takesue}.
The Hamiltonian of many--particle system is given by
\begin{eqnarray}
	{\cal H} = \sum_i \frac{\left.\vc{p}_i\right.^2}{2m_i} + U(\{\vc{x}\}),
	\label{eq:j6_2_1}
\end{eqnarray}
where $\vc{p}_i$ and $m_i$ are the $i$--th particle momentum and mass,
$U(\{\vc{x}\})$ is the total interaction potential which depends on plural particle positions $\{\vc{x}\}$.

First, we consider the system of particles.
The $i$--th particle energy $e_i(t)$ is
 localized in the $i$--th particle position $\vc{x}_i$.
Thereby, the microscopic energy field $e(\vc{x},t)$ is defined by
\begin{eqnarray}
	e(\vc{x},t)=\sum_i e_i(t) \delta(\vc{x}-\vc{x}_i(t)),
	\label{eq:j6_2_2}
\end{eqnarray}
where $\delta(\vc{x})$ is Dirac's delta function in three dimensional space as 
\begin{eqnarray}
	\delta(\vc{x}) = \delta(x)\delta(y)\delta(z),
	\label{eq:j6_2_3}
\end{eqnarray}
and
\begin{eqnarray}
	\vc{x} = \rbk{\mx{x \\ y \\ z \\}}.
	\label{eq:j6_2_4}
\end{eqnarray}

The $i$--th particle kinetic energy ${\left.\vc{p}_i\right.^2}/{2m_i}$
 is possessed by the $i$--th particle energy $e_i(t)$,
 and is not possessed by $e_{k}(t),(k \neq i)$.
Moreover, we assume that 
 the total interaction energy $U(\{\vc{x}\})$ 
 is allocated into particles
 and is localized only in their positions.
The interaction energy quota of the $i$--th particle is described by $u_i$.
Therefore,
\begin{eqnarray}
	e_i(t) = \frac{\vc{p}_i(t)^2}{2m_i} + u_i(t).
	\label{eq:j6_2_5}
\end{eqnarray}
A point to notice is that
 the interaction energy quota $u_i(t)$ is merely an artificial quantity.
The differential of the interaction energy quota $u_i(t)$ cannot
 generate forces acting on particles.
The differential of the total interaction potential $U(\{\vc{x}\})$ corresponds to the force.
A main problem in the present paper is to determine the interaction energy quota $u_i(t)$.
We describe the method to solve this problem in section 2 and 3.

Because the space integration of the microscopic energy field $e(\vc{x},t)$
 is equal to the total energy $E$ in the system,
 we obtain the following condition for the $i$--th particle energy $e_i(t)$,
\begin{eqnarray}
	E = \int dx^3 e(\vc{x},t) = \sum_i e_i(t).
	\label{eq:j6_2_6}
\end{eqnarray}
The total energy $E$ obeys energy conservation law.


If microscopic energy current field $\vc{j}(\vc{x},t)$ exists,
 it satisfies the following `continuity equation' with 
 the microscopic energy field $e(\vc{x},t)$;
\begin{eqnarray}
	\pdif{e(\vc{x},t)}{t} + \nabla \cdot \vc{j}(\vc{x},t) = 0.
	\label{eq:j6_2_7}
\end{eqnarray}
The time derivative of the $e(\vc{x},t)$ becomes
\begin{eqnarray}
	\pdif{e(\vc{x},t)}{t} &=& \sum_i \dif{e_i(t)}{t} \delta(\vc{x}-\vc{x}_i(t)) \retn
		&&- \sum_i e_i(t) \dot{\vc{x}}_i(t) \cdot \nabla\delta(\vc{x}-\vc{x}_i(t)),
		\indent \label{eq:j6_2_8}
\end{eqnarray}
where, in three dimensional space, $\nabla\delta(\vc{x})$
 is a column vector as 
\begin{eqnarray}
	\nabla\delta(\vc{x}) = \rbk{\mx{\pdif{}{x}\delta(x)\delta(y)\delta(z)\smallskip \\
					\delta(x)\pdif{}{y}\delta(y)\delta(z)\smallskip \\
					\delta(x)\delta(y)\pdif{}{z}\delta(z)\smallskip \\ }}.
		\label{eq:j6_2_9}
\end{eqnarray}

When the $i$--th particle moves with velocity $\dot{\vc{x}}_i$,
 the energy current $e_i \dot{\vc{x}}_i$ is generated and
 is localized only in the $i$--th particle position $\vc{x}_i$
 because the energy is localized only in particle positions as Eq.(\ref{eq:j6_2_2}).
Moreover, the energy current between particles is generated 
 by the interaction between particles.
From these point of view,
 we propose the microscopic energy current field $\vc{j}(\vc{x},t)$ as follow;
\begin{eqnarray}
	\vc{j}(\vc{x},t)&=& \sum_i e_i(t) \dot{\vc{x}}_i(t) \delta(\vc{x}-\vc{x}_i(t)) \retn
	 && + \sum_{i,k>i} j_{i \ra k}(t) \frac{\vc{\xi}_{ki}}{\abs{\vc{\xi}_{ki}}}
		\delta(\vc{x};\mbox{segment},i \ra k),\retn
	\label{eq:j6_2_10}
\end{eqnarray}
where $j_{i \ra k}$ is the magnitude of the energy current from the $i$--th particle
 to the $k$--th particle, 
 $\vc{\xi}_{ki}$ is relative position vector as 
\begin{eqnarray}
	\vc{\xi}_{ki} \equiv \vc{x}_k - \vc{x}_i.
	\label{eq:j6_2_11}
\end{eqnarray}
The localization function like the line segment from the $i$--th particle
 to the $k$--th particle is defined by 
\begin{eqnarray}
	&&\delta(\vc{x};\mbox{segment},i \ra k) \retn
	 &&\equiv (\vc{\xi}_{ki})_x \abs{\vc{\xi}_{ki}}
		\delta\Big((\vc{\xi}_{ki}\times (\vc{x} - \vc{x}_i))_y\Big) \retn
	&&\indent \times \delta\Big((\vc{\xi}_{ki}\times (\vc{x} - \vc{x}_i))_z\Big) \retn
	&&\indent \times \theta\Big(\vc{\xi}_{ki}\cdot \rbk{\vc{x} - \vc{x}_i}\Big)
		\theta\Big(\vc{\xi}_{ik}\cdot \rbk{\vc{x} - \vc{x}_k}\Big),\retn
	\label{eq:j6_2_12}
\end{eqnarray}
where $(\cdots)_y$ and $(\cdots)_z$ are y-- and z--components of a vector,
 and $\theta$ is the Heaviside step function.
The first term of the right--hand in Eq.(\ref{eq:j6_2_10}) means the energy current due to the movement of particles,
 and the second term is the energy current between particles owing to the interaction of particles.

The localization function $\delta(\vc{x};\mbox{segment},i \ra k)$ has following property;
\begin{eqnarray}
	 &&\frac{\vc{\xi}_{ki}}{\abs{\vc{\xi}_{ki}}} \nabla \cdot \delta(\vc{x};\mbox{segment},i \ra k)\retn
	&&\indent= \delta(\vc{x}-\vc{x}_i) - \delta(\vc{x}-\vc{x}_k).\retn
	\label{eq:j6_2_13}
\end{eqnarray}
From  Eqs.(\ref{eq:j6_2_10}) and (\ref{eq:j6_2_13}), 
 the divergence of the microscopic energy current field $\vc{j}(\vc{x},t)$
 becomes
\begin{eqnarray}
	\nabla \cdot \vc{j}(\vc{x},t) &=& \sum_i e_i(t) \dot{\vc{x}}_i(t) \cdot \nabla \delta(\vc{x}-\vc{x}_i(t)) \retn
	&&+ \sum_{i,k\neq i} j_{i \ra k}(t) \delta(\vc{x} - \vc{x}_i(t)),
	\indent \label{eq:j6_2_14}
\end{eqnarray}
where $j_{k \ra i} = - j_{i \ra k}$.
The function $\delta(\vc{x} - \vc{x}_i(t))$ appears
 in second term of Eq. (\ref{eq:j6_2_14}),
 because the energy current between particle exists
 only on the line segment between particles in (\ref{eq:j6_2_10}).
By comparison between Eqs. (\ref{eq:j6_2_8}), (\ref{eq:j6_2_14}),
 we obtain the condition for the energy current between particles
\begin{eqnarray}
	\dif{e_i(t)}{t} = - \sum_{k \neq i} j_{i \ra k}(t).
	\label{eq:j6_2_15}
\end{eqnarray}
Thus, when the time derivative of the $i$--th particle energy $e_i(t)$ can be
 composed by summation $\displaystyle \sum_{k \neq i}$, 
 we can regard the elements of the summation as the energy current between particles $j_{i \ra k}$.
Because the $i$--th particle energy $e_i(t)$ is defined by Eq. (\ref{eq:j6_2_5}),
We must determine the interaction energy quota $u_i(t)$
 to obtain the energy current between particles $j_{i \ra k}$.


\section{Energy Current with\\Two--Body Force}

We explain the interaction energy quota $u_i(t)$ according to two--body interaction.
As a result, we obtain the energy current between particles $j_{i \ra k}$.
The two--body interaction potential function
 is described by $\phi_{ik}(r_{ik})$, where $r_{ik} \equiv \abs{\vc{x}_i-\vc{x}_k}$.
We set the following localization with the constants $a_i^{(ik)}$ and $a_i^{(ik)}$;
 the interaction energy localized in $i$--th particle position
 is $a_i^{(ik)}\phi_{ik}(r_{ik})$,
 and the interaction energy localized in $k$--th particle position
 is $a_k^{(ik)}\phi_{ik}(r_{ik})$.
Thereby, the interaction energy quota $u_i$ was defined by
\begin{eqnarray}
	u_i(t) \equiv \sum_{k_\neq i} {}' a_i^{(ik)}\phi_{ik}(r_{ik}),
	\label{eq:j6_3_1}
\end{eqnarray}
where
 $\displaystyle \sum_{k_\neq i} {}'$ means summation about particles
 doing interaction with the $i$--th one.

From the right--hand of Eq. (\ref{eq:j6_3_1}),
 the interaction energy quota $u_i$ depends 
 only on canonical variables.
The kinetic energy ${\left.\vc{p}_i\right.^2}/{2m_i}$
 is also composed only of canonical variables.
Therefore, the $i$--th particle energy $e_i$ is the function only of canonical variables, 
 the Liouville operator can be applied to
 the time derivative of the $i$--th particle energy $e_i$ as
\begin{eqnarray}
	\dif{e_i}{t} = - \brc{H, e_i}
	 = - \sum_{k \neq i}\brc{e_k, e_i}.
	\label{eq:j6_3_2}
\end{eqnarray}
By comparison between Eqs. (\ref{eq:j6_2_15}) and (\ref{eq:j6_3_2}),
 the energy current from the $i$--th particle to the $k$--th one $j_{i \ra k}$
 is determined as the element of the summation in Eq. (\ref{eq:j6_3_2}),
\begin{eqnarray}
	j_{i \ra k} &=& \brc{e_k, e_i} \retn
		&=& \frac{\vc{p}_i}{2 m_i}\cdot \pdif{u_k}{\vc{x}_i}
		 - \frac{\vc{p}_k}{2 m_k}\cdot \pdif{u_i}{\vc{x}_k} \retn
		&=& \rbk{a_k^{(ik)}\frac{\vc{p}_i}{m_i} + a_i^{(ik)}\frac{\vc{p}_k}{m_k}}
			\cdot \pdif{\phi_{ik}(r_{ik})}{\vc{x}_i}.
		\indent \label{eq:j6_3_3}
%
\end{eqnarray}

To explain the constants $a_i^{(ik)}$ and $a_k^{(ik)}$, let us consider the system of the two particles.
The difference of the 1-st particle energy $e_1(t)$ between the times $t_1$ and $t_2$ is the follows
 from Eqs. (\ref{eq:j6_2_5}) and (\ref{eq:j6_3_1});
\begin{eqnarray}
	e_1(t_2) - e_1(t_1) = \frac{\vc{p}_i(t_2)^2}{2m_i} - \frac{\vc{p}_i(t_1)^2}{2m_i} \retn
		 \indent + a_1^{(12)}\sbk{\phi_{12}(r_{12}(t_2)) - \phi_{12}(r_{12}(t_1))}.
		\label{eq:j6_3_4}
\end{eqnarray}
On the other hand, the difference of the 1-st particle energy $e_1(t)$ between the times $t_1$ and $t_2$
 can be described from Eqs. (\ref{eq:j6_2_15}) and (\ref{eq:j6_3_3}) by
\begin{eqnarray}
	e_1(t_2) - e_1(t_1)
	= - \int_{t_1}^{t_2} j_{1 \ra 2} dt \retn 
	\indent= -\int_{t_1}^{t_2} \rbk{a_1^{(12)} + a_2^{(12)}} \frac{\vc{p}_1}{m_1} \cdot \pdif{\phi_{12}(r_{12})}{\vc{x}_1} dt\retn
	\indent \hspace{1em} + \int_{t_1}^{t_2} a_1^{(12)} \rbk{\dot{\vc{x}}_1 - \dot{\vc{x}}_2} \cdot \pdif{\phi_{12}(r_{12})}{(\vc{x}_1 - \vc{x}_2)} dt \retn 
	\indent= \rbk{a_1^{(12)} + a_2^{(12)}}\rbk{\frac{\vc{p}_i(t_2)^2}{2m_i} - \frac{\vc{p}_i(t_1)^2}{2m_i}} \retn
	\indent \hspace{1em} + a_1^{(12)}\sbk{\phi_{12}(r_{12}(t_2)) - \phi_{12}(r_{12}(t_1))}.
		\label{eq:j6_3_5}
\end{eqnarray}
For Eqs. (\ref{eq:j6_3_4}) and (\ref{eq:j6_3_5}), the constants satisfy
\begin{eqnarray}
 a_i^{(ik)} + a_k^{(ik)} = 1.
		\label{eq:j6_3_6}
\end{eqnarray}

Because the both of two particles
 possess just half of the two--body interaction energy historically,
 the constants are determined by
\begin{eqnarray}
	a_i^{(ik)} = a_k^{(ik)} = \frac{1}{2}.
		\label{eq:j6_3_7}
\end{eqnarray}



\section{Energy Current with\\Multi--Body Force}

We assume that the interaction energy was localized in particle positions
 even if the multi--body force appears.
We abandon that the interaction energy quota $u_i$ is directly proportionate to the multi--body interaction energy
 as is the case with the two--body interaction.
We propose new method to determine the interaction energy quota $u_i$.

First, we explain the multi--body interaction potential.
The total interaction potential $U(\{\vc{x}\})$
 is the sum of the multi--body interaction potential.
It can include the two--body interaction potential.
However, it does not include the on--site potential.
The necessary and sufficient condition to satisfy the total momentum conservation
 is that the multi--body interaction potential depends only on
 relative position vectors between particles $\{\vc{\xi}\}$ as
\begin{eqnarray}
	 U(\{\vc{x}\}) = U(\{\vc{\xi}\}).
	\label{eq:j6_4_1}
\end{eqnarray}
The multi--body force acting on the $i$--th particle due to the variation of the relative position vector
 between the $i$--th particle and the $k$--th one $\vc{\xi}_{ik}$ is described by
 $\vc{F}_i^{(ik)}$,
 and has the following property;
\begin{eqnarray}
	\vc{F}_i^{(ik)} &=& - \pdif{U(\{\vc{\xi}\})}{\vc{\xi}_{ik}} 
		\pdif{\vc{\xi}_{ik}}{\vc{x}_i} \retn
		&=& \pdif{U(\{\vc{\xi}\})}{\vc{\xi}_{ki}}\pdif{\vc{\xi}_{ki}}{\vc{x}_k} 
		= - \vc{F}_k^{(ki)}. 
	\label{eq:j6_4_2}
\end{eqnarray}
It is namely that the forces acting on the $i$--th and $k$--th particles and the  one with the variation of $\vc{\xi}_{ik}$
 are equal in magnitude and opposite in direction.
This is the law of the multi--body force corresponds to
 Newton's third law (the law of action and reaction).
The total multi--body force acting on the $i$--th particle $\vc{F}_i$ is given by 
\begin{eqnarray}
	\vc{F}_i = - \pdif{U(\{\vc{\xi}\})}{\vc{x}_i} = \sum_{k \neq i} \vc{F}_i^{(ik)}.
	\label{eq:j6_4_3}
\end{eqnarray}
Therefore, the Eq. (\ref{eq:j6_4_2}) implies the total momentum conservation as
\begin{eqnarray}
	\dif{}{t}\sum_i \vc{p}_i &=& \sum_i \vc{F}_i 
	 = \sum_i \sum_{k \neq i} \vc{F}_i^{(ik)} \retn
	 &=& \sum_i \sum_{k > i} \rbk{\vc{F}_i^{(ik)} + \vc{F}_k^{(ki)} } \retn
	 &=& 0.
	\label{eq:j6_4_4}
\end{eqnarray}

We consider a virtual frame which moves at a velocity $\vc{\beta}$
 relative to the original frame as
\begin{eqnarray}
	\vc{\tilde{x}}_i(t) = \vc{x}_i(t) - \int^t \vc{\beta}(t')dt',
	\label{eq:j6_4_5}
\end{eqnarray}
where quantities in the virtual frame are marked tilde as 
We define the interaction energy quota $u_i$ as the negative work
 of the force acting on the $i$--th particle until time $t$
 in the virtual frame
by
\begin{eqnarray}
	u_i(t) &\equiv& - \int^t \vc{\tilde{F}}_i(t') \cdot \dot{\vc{\tilde{x}}}_i(t') dt',
	\label{eq:j6_4_7}
\end{eqnarray}
where $\vc{\tilde{F}}_i$ is the multi--body force acting on the $i$--th particle
 in the virtual frame as
\begin{eqnarray}
	\vc{\tilde{F}}_i = - \pdif{U(\{\vc{\tilde{\xi}}\})}{\vc{\tilde{x}}_i}.
	\label{eq:j6_4_8}
\end{eqnarray}

Relative position vectors and their derivatives are
 invariant under the virtual frame as
\begin{eqnarray}
	\vc{\tilde{\xi}}_{ik} = \vc{\xi}_{ik},\label{eq:j6_4_9} \\
	\pdif{\vc{\tilde{\xi}}_{ik}}{\vc{\tilde{x}}_l}
	 = \pdif{\vc{\xi}_{ik}}{\vc{x}_l}
	 = \delta_{il} - \delta_{kl}.
	\label{eq:j6_4_10}
\end{eqnarray}
Because the total interaction potential $U(\{\vc{\xi}\})$ 
depends only on the relative position vectors, 
it is invariant under the virtual frame as
\begin{eqnarray}
	U(\{\vc{\tilde{\xi}}\})
	 = U(\{\vc{\xi}\}).
	\label{eq:j6_4_11}
\end{eqnarray}
Thereby, the multi--body forces are also invariant under the virtual frame as
\begin{eqnarray}
\vc{\tilde{F}}_i^{(ik)}
	= - \pdif{U(\{\vc{\tilde{\xi}}\})}{\vc{\tilde{\xi}}_{ik}}
	= - \pdif{U(\{\vc{\xi}\})}{\vc{\xi}_{ik}}
	= \vc{F}_i^{(ik)},\retn
	\label{eq:j6_4_13}\\
	\vc{\tilde{F}}_i = \sum_{k \neq i} \vc{\tilde{F}}_i^{(ik)}
	= \sum_{k \neq i} \vc{F}_i^{(ik)}
	= \vc{F}_i.
	\label{eq:j6_4_14}
\end{eqnarray}

For Eqs. (\ref{eq:j6_4_5}) and (\ref{eq:j6_4_14}),
 we can describe the interaction energy quota $u_i$ of Eq. (\ref{eq:j6_4_7})
 by the quantities in the original frame as
\begin{eqnarray}
	u_i(t) = 
		- \int^t \vc{F}_i(t') \cdot \rbk{\dot{\vc{x}_i}(t')-\vc{\beta}(t')} dt'.
	 \indent \label{eq:j6_4_15}
\end{eqnarray}
From Eqs. (\ref{eq:j6_2_5}), (\ref{eq:j6_4_15}) and canonical equations of motion,
 the time derivative of the $i$--th particle energy $e_i$ become
\begin{eqnarray}
	\dif{e_i}{t} = \dot{\vc{p}_i} \cdot \frac{\vc{p}_i}{m_i}
	- \vc{F}_i \cdot \dot{\vc{x}_i}
	  + \vc{F}_i \cdot \vc{\beta}
	= \vc{F}_i \cdot \vc{\beta}.
	\hspace{2em} \label{eq:j6_4_16}
\end{eqnarray}
This implies that the energy current between particles $j_{i \ra k}$
 is generated in spite of satisfaction with 
the energy conservation law Eq. (\ref{eq:j6_2_6})
 from Eqs.(\ref{eq:j6_4_2}) and (\ref{eq:j6_4_3}) as
\begin{eqnarray}
	\dif{}{t} \sum_i e_i
	 &=& \sum_i \sum_{k > i} \rbk{\vc{F}_i^{(ik)} + \vc{F}_k^{(ki)}}
			\cdot \vc{\beta} \retn
	 &=& 0.
	\label{eq:j6_4_17}
\end{eqnarray}

Moreover, the plural virtual frame
 can be employed
 as long as the energy conservation Eq. (\ref{eq:j6_2_6}) is satisfied.
We apply different velocity $\vc{\beta}_{ik}$
 in each force due to the variation of the relative position vector $\vc{\xi}_{ik}$.
The interaction energy quota $u_i(t)$ is conclusively defined by
\begin{eqnarray}
	u_i(t) \equiv
		- \int^t \sum_{k \neq i}  \vc{F}_i^{(ik)}(t')
		\cdot \rbk{\dot{\vc{x}_i}(t')-\vc{\beta}_{ik}(t')} dt',\retn
	\label{eq:j6_4_18}
\end{eqnarray}
where Eqs. (\ref{eq:j6_4_5}) and (\ref{eq:j6_4_13}) are used.
This new definition (\ref{eq:j6_4_18}) implies the change of 
 the time derivative of the $i$--th particle energy $e_i(t)$ 
 from Eq. (\ref{eq:j6_4_16}) to
\begin{eqnarray}
	\dif{e_i}{t} = \sum_{k \neq i} \vc{F}_i^{(ik)} \cdot \vc{\beta}_{ik}.
	\label{eq:j6_4_19}
\end{eqnarray}
For the energy conservation law (\ref{eq:j6_2_6}) and Eq. (\ref{eq:j6_4_2}) give 
 the following condition to determine the velocity $\vc{\beta}_{ik}$;
\begin{eqnarray}
	\dif{}{t}\sum_i e_i &=& \sum_i \sum_{k \neq i} \vc{F}_i^{(ik)} \cdot \vc{\beta}_{ik}\retn
	&=&\sum_i \sum_{k > i}
		\rbk{ \vc{F}_i^{(ik)} \cdot \vc{\beta}_{ik}
		+ \vc{F}_k^{(ki)} \cdot \vc{\beta}_{ki}}\retn
	&=& \sum_i \sum_{k > i}
		\vc{F}_i^{(ik)} \cdot \rbk{\vc{\beta}_{ik} - \vc{\beta}_{ki}} \retn
	 &=& 0.
	\label{eq:j6_4_20}
\end{eqnarray}
This condition (\ref{eq:j6_4_20}) implies
 that the velocity $\vc{\beta}_{ik}$ must be independent of the order 
 of the subscripts $i$ and $k$.
The velocity $\vc{\beta}_{ik}$ has the dimension of the velocity,
 and depends on only the $i$--th and $k$--th particles.
Therefore, we define 
\begin{eqnarray}
	\vc{\beta}_{ik} = \vc{\beta}_{ki} \equiv c_i \dot{\vc{x}}_i + c_k \dot{\vc{x}}_k,
	\label{eq:j6_4_21}
\end{eqnarray}
where the coefficients $c_i$ and $c_k$ are constants.

By comparison between Eqs. (\ref{eq:j6_2_15}) and (\ref{eq:j6_4_19}),
 the energy current from the $i$--th particle to the $k$--th one is determined by
\begin{eqnarray}
	j_{i \ra k} \equiv - \vc{F}_i^{(ik)} \cdot
		 \rbk{ c_i \dot{\vc{x}}_i + c_k \dot{\vc{x}}_k}.
	\label{eq:j6_4_22}
\end{eqnarray}

Moreover, when the total interaction potential $U(\{\vc{\xi}\})$ is composed of the two--body interaction,
 the energy current from the $i$--th particle to the $k$--th one $j_{i \ra k}$ (\ref{eq:j6_4_22})
 should be consistent with Eq. (\ref{eq:j6_3_3}).
Therefore, all of the constants $c_i$ and $c_k$
 are determined by
\begin{eqnarray}
	c_i = a_k^{(ik)} = \frac{1}{2}, \indent
	c_k = a_i^{(ik)} = \frac{1}{2}.
	\label{eq:j6_4_23}
\end{eqnarray}
%
The energy current from the $i$--th particle to the $k$--th one $j_{i \rightarrow k}$
 is given by
\begin{eqnarray}
	j_{i \rightarrow k} \equiv 
		 -\frac{1}{2}\rbk{ \dot{\vc{x}}_i + \dot{\vc{x}}_k}
		 \cdot \vc{F}_i^{ik}.
	\label{eq:j6_4_24}
\end{eqnarray}

\section{Discussion}

In present work,
 we extended the theory of the microscopic energy field and energy current 
 to the system of the multi--body force.
The $i$--th particle interaction energy quota $u_i$ is defined by Eq. (\ref{eq:j6_4_18})
 as the negative work of the forces acting on the $i$--th particle until time $t$
 in the virtual frames moving at the velocity $\vc{\beta}_{ik}$ with each variation
 of relative position vector $\vc{\xi}_{ik}$.
As a result, the energy current between particles $j_{i \rightarrow k}$ is obtained.
%
%

%

In the one particle system, the interaction potential is the negative work on the particle.
By analogy with this, at first
we defined the interaction energy quota $u_i$ as the negative work
  of the force acting on the $i$--th particle until time $t$ in the original frame by
\begin{eqnarray}
	u_i(t) \equiv - \int^t \vc{F}_i(t') \cdot \dot{\vc{x}_i}(t') dt',
	\label{eq:j6_5_1}
\end{eqnarray}
where $\vc{F}_i(t)$ is the force acting on the $i$--th particle Eq. (\ref{eq:j6_4_3}).
%
From Eqs. (\ref{eq:j6_2_5}), (\ref{eq:j6_5_1}) and canonical equations of motion,
 the time derivative of the $i$--th particle energy $e_i$ become
\begin{eqnarray}
	\dif{e_i}{t} = \dot{\vc{p}_i} \cdot \frac{\vc{p}_i}{m_i}
	 	- \vc{F}_i \cdot \dot{\vc{x}_i}
	 = 0.
	\label{eq:j6_5_2}
\end{eqnarray}
Therefore, for the Eq. (\ref{eq:j6_2_15}),
 the energy current between particles $j_{i \rightarrow k}$ becomes zero.
The microscopic energy current field $\vc{j}(\vc{x},t)$
 is namely according to only movement of particles.

Now, in the macroscopic fluid dynamics,
 the energy current includes a stress term and a heat current term
 besides the term of the movement of a fluid particle.
If the microscopic energy field should be connected with the macroscopic field,
 the energy current between particles $j_{i \rightarrow k}$ need to be not zero.
The definition (\ref{eq:j6_5_1}) is regard as a failure.

We notice that the multi--body interaction potential and its forces are invariant
 under the transformation of a reference frame.
The virtual frame moving at velocity $\vc{\beta}_{ik}$ relative to the original frame
 is used.
As a result, the energy current between particles is generated.
Although the time derivative of the $i$--th particle energy $e_i(t)$ (\ref{eq:j6_4_19})
 becomes not zero, 
 the energy conservation law is satisfied 
 owing to the law of the multi--body force (\ref{eq:j6_4_2})
 corresponding to Newton's third law as Eq. (\ref{eq:j6_4_20}).


A problem is the determination of the constants $a_i^{(ik)}$, $a_k^{(ik)}$, $c_i$ and $c_k$.
When the interaction is due to the two--body force,
 it is suitable to split the two--body interaction energy into two particles just half
 as Eq.(\ref{eq:j6_3_7}).
The cause of the splitting half is not that two particles has same properties
 like a charge, a mass, a sort of atom, and so on.
The cause must be 
that the forces on two particles are equal in magnitude due to Newton's third law.
The constants $c_i$ and $c_k$ is also determined by $1/2$ as Eq. (\ref{eq:j6_4_23}) 
 due to the law of the multi--body force (\ref{eq:j6_4_2}) corresponding to Newton's third law.


%
%
%

%
%
%

Our method for the multi--body force yields new information $j_{i \ra k}$ to the CMD.
However, couples of the $i$--th and $k$--th particles for $j_{i \ra k}$ depend on
 the selection of the independent variables $\{\vc{\xi}\}$
 of the interaction potential $U(\{\vc{\xi}\})$.
In the CMD under the multi--body force,
 we consider that the independent variables $\{\vc{\xi}\}$ are chosen
 as covalent bonds.

%
%

\section{Conclusion}

The theory of the microscopic energy field and the energy current was used with only two--body force.
We propose the new method to deal with the multi--body force in the present paper.
The interaction energy quota $u_i$ in Eq. (\ref{eq:j6_2_5}) is defined by Eq. (\ref{eq:j6_4_18}).
It means the negative work of the forces acting on the $i$--th particle until time $t$
 in the virtual frames moving at the velocity $\vc{\beta}_{ik}$ with each variation
 of relative position vector $\vc{\xi}_{ik}$.
Then, the velocity $\vc{\beta}_{ik}$ is determined by Eq. (\ref{eq:j6_4_21}).
Our method is based on the law of the multi--body force Eq. (\ref{eq:j6_4_2})
 corresponding to Newton's third law.
As a result, the theory of the microscopic energy field and the energy current is extended
 to the multi--body force.
The energy current between particles $j_{i \ra k}$ is derived as Eq. (\ref{eq:j6_4_24}).

\section*{Acknowledgments}

The authors thank Dr. Shinji Takesue and Dr. Akira Ueda for helpful comments.
The work is supported partly  by the National Institutes of Natural Sciences
 undertaking for Forming Bases for Interdisciplinary and International Research
 through Cooperation Across Fields of Study, and partly by Grand-in Aid for
 Exploratory Research (C), 2006, No.~17540384 from the Ministry of Education,
 Culture, Sports, Science and Technology.


\end{sloppypar}
\end{document}